\documentclass[12pt]{iopart}
\usepackage{graphicx}
\usepackage{dcolumn}
\usepackage{color}
\usepackage{graphics,graphicx}
\usepackage{psfrag}
\usepackage{graphics,graphicx}
\usepackage{dcolumn,bm}
\newcommand{\cu}
{\address{Department of Physics, University of Calcutta,
92 Acharya Prafulla Chandra Road, Kolkata 700009, India.}}
\newcommand{\tifr}
{\address{
Department of Theoretical Physics, Tata Institute of Fundamental Research, Homi Bhabha Road, Mumbai 400 005, India.}}

\begin{document}

\title{Universal features of exit probability in opinion dynamics models with domain size dependent dynamics}

\author{Parna Roy}%
\cu
\author{Soham Biswas}%
\tifr
\author{Parongama Sen}%
\cu

\begin{abstract}

We study the exit probability for several binary opinion dynamics
models in one dimension in which the opinion state (represented by $\pm 1$) 
of an agent is determined 
by  dynamical rules dependent on the  size of its neighbouring 
domains. In all these models, we find the 
exit probability behaves like a step function in the thermodynamic limit.
In a finite system of size $L$, the  exit probability $E(x)$ as a function of the initial 
fraction $x$ of one type of opinion is given by  $E(x) = f[(x-x_c)L^{1/\nu}]$
with a universal value of $\nu = 2.5 \pm 0.03$.
The form of the scaling function is also universal:  $f(y) = [\tanh(\lambda y +c) +1]/2$, 
 where  $\lambda$ is found to be dependent on the particular dynamics. 
The variation of $\lambda$ against the parameters of the models is studied.  
$c$ is non-zero only when the  dynamical rule  distinguishes between
$\pm 1$ states; comparison with theoretical estimates in this case shows very good agreement.

\end{abstract}


\maketitle

\section{Introduction}

Identifying universality in physical phenomena occurring in 
different systems has become an  important topic of  research in the last
few decades. 
Universality usually indicates that there are some fundamental common underlying  
features in the systems under consideration.
 It also signifies that there are only a few  
  parameters occurring in the 
systems those are relevant. Existence of universality  justifies the 
study of models which include only these  parameters while  real 
systems  are far more complicated.  
Universality has been observed in critical phenomena;
for example  there is  a unique value of 
the  order parameter exponent in liquid-gas phase transition \cite{skma}. 
 Universal features may also 
appear away from criticality as for example the 
universal scaling behaviour obtained for characteristic 
features of many complex networks \cite{socio}.
In dynamical phenomena, universal classes  have been 
observed close to and away from criticality.  
Models belonging to the same static universality class  
may belong to a different universality class as far as dynamics
is concerned \cite{hoha}. Non-equilibrium dynamics  
also reveal dynamical universal classes, for example, 
many systems have been shown  to belong to the directed percolation universality class \cite{Odor,Hinrichsen}.

Exit probability (EP) is one important feature of  dynamical models with two absorbing states. Examples include binary opinion dynamics models and Ising spin models.
Exit probability $E(x)$ denotes the probability that the system ends up with 
all opinion/spins in a certain state when initially $x$ fraction had been
in that state. 
Recently, a lot of effort has been put to identify universal features of exit probability in opinion dynamics models (as well as generalised Ising-Glauber models)
in one dimension.  In the voter model  (which is equivalent to the  Ising-Glauber dynamics in one-dimension) $E(x)$ is simply equal to $x$ while
 $E(x)$ is a non-linear continuous function of $x$ in nonlinear voter model,  Sznajd model and long ranged Ising Glauber model  \cite{Castellano,Slanina,Lambiotte,parna1} in one dimension.
 In all these cases $E(x)$ apparently shows no dependence on finite sizes for larger system sizes \cite{Castellano}.
 The exit probability was also calculated and generalised for nonlinear q-voter model in one-dimension \cite{qvoter,timparno,new}.
In higher dimensions or on networks, the exit probability 
in the thermodynamic limit may exhibit a step function behaviour, interpreted 
as a phase transition in some earlier works \cite{stauffer,networks,step1,step2,step3}. Strong finite size effects are observed here.
The  possibility of problems arising while calculating the exit probability, due to the sole use of local update rules in dynamical systems, was 
addressed previously by Galam et al \cite{galamp}.

In a recent study of a binary opinion dynamics model \cite{bss}, the behaviour 
of the exit probability  was found to be quite different from the well studied 
models mentioned in the paragraph above; it exhibited a step function like behaviour in the thermodynamic limit {\it even in one dimension}.
It also showed the existence of an exponent with a value independent of the model parameter.
In the model considered in \cite{bss}, the  state (spin or opinion) of
an agent was updated according to a rule dependent on the size of his/her two neighbouring domains.

 In this paper, we investigate whether the step function behaviour of $E(x)$ in the thermodynamic limit
is a universal  feature of 
models with dynamical rules which involve the sizes of neighbouring domains in one dimension.
Careful study of a number of models indicates that indeed such a universal feature exists. Furthermore, universal scaling function and 
an exponent with model independent universal value are obtained. The non-universal quantities associated with the scaling function 
also show very interesting behaviour as a function of the model parameters. 

In section II, the models are introduced  and details of the  simulation provided briefly. Results are presented in section III and discussions and summary in the last section.

\section{The Models}

We have considered a number of models which  mimic opinion formation 
in a society where the opinions have values $\pm 1$.
These states can be  equivalently regarded as the states of Ising spins
and the models may be interpreted  as interacting spin models as well.
Since it is convenient  to talk in terms of spins 
we will use the term spin instead of 
opinion henceforth.
It also becomes more meaningful as the models 
behave as familiar spin models in certain limits. 

In all these  models considered in one dimension, in the spin picture, 
the
spins located on the domain boundaries are liable to flip, as in the 
case of zero temperature Ising model with Glauber dynamics. The spins which can undergo 
change have therefore two neigbouring domains of opposite spin states.
In general, in the models considered in this work,    
the state of the  spin at the domain boundary is determined by the state
of the neighbouring domains and their sizes.  

\subsubsection{BS model}
In the first model introduced in this class by Biswas and Sen \cite{biswas-sen1}, the BS model
hereafter, the state of the spin simply follows that of the larger 
neighbouring domain. 
Hence if $d_{up}$ and $d_{down}$ are the neighbouring domain sizes (with up and down spins respectively), the spin will be up if 
$d_{up} > d_{down}$ and down otherwise.
In case   $d_{up} = d_{down}$,
 the state is chosen to be $\pm 1$ with equal probability.
A spin sandwiched between domains of opposite sign is always flipped.  
The BS model, where the final configuration is all up or all down states,
is different from the Ising model having different dynamical exponents
with respect to domain growth and persistence \cite{biswas-sen1}.
\subsubsection{BS model with cutoff}
In the BS model one can introduce a cutoff \cite{biswas-sen2} 
on the  size of the domain 
while calculating $d_{up}$ and $d_{down}$.
The cutoff is taken as  $R=pL/2$ where $L$ is the system size and $p$ a parameter ranging from zero to 1. 
Now, with the introduction of this cut off parameter $p$,  the definition of $d_{up}$ and $d_{down}$ are
modified: $d_{up} = {\rm{min}}\{R, d_{up}\}$ and similarly  $d_{down} = {\rm{min}}\{R, d_{down}\}$ while the same dynamical rule explained earlier applies. 
When $p$ is infinitesimal, the results are identical to  those of the nearest neighbour Ising model. For finite 
$p$, there is a crossover behaviour in time: initially there is a BS-like behaviour after which very few domain walls survive which perform almost noninteracting motion for a long time before annihilating each other. 
$p=1$ is of course equivalent to the BS model.  
Instead of a fixed value of the cutoff a  random cutoff can also be considered.
\subsubsection{The $\beta$ model}
We have also studied a  model  where the dynamics 
depends on the size of the neighbouring domains stochastically 
with a noise like parameter $\beta$ \cite{sen}.
In this so called $\beta$ model, 
the   probability of a boundary spin to be up  is taken as  
\begin{equation}
P(up) \propto e^{\beta(d_{up}- d_{down})},
\end{equation}
and it is down with probability 
\begin{equation}
P(down) \propto e^{\beta(d_{down}- d_{up})}.
\end{equation}
The normalised probabilities are  therefore 
$P(up) =  \exp{\beta\Delta}/(\exp(\beta\Delta) + \exp(-\beta\Delta))$ and  
 $P(down)=1-P(up)$,
where $\Delta = (d_{up}- d_{down}) $. 
$\beta =0$ is equivalent to the Ising model and any finite value of $\beta$ drives the system to the BS dynamical class. 

\subsubsection{The $\epsilon$ model}
The BS model was shown to be equivalent to a reaction diffusion model in one dimension 
where random walkers tend to walk towards their nearer neighbours 
and annihilate on meeting \cite{bsr}. This reaction diffusion  model can be generalised 
by assigning a  probability $\epsilon$ to move towards the nearer neighbour.
 $\epsilon = 1$ corresponds to the BS model and $\epsilon = 0.5$, the model
with unbiased  walkers which mimics the coarsening dynamics in  the 
 Ising model. 
We call this model the $\epsilon$ model. In the equivalent spin model,  
the larger neighbouring domain will dictate the sign of the spin 
on the boundary with probability $\epsilon$. 
\subsubsection{The $\rho-\mu$ model}
Another stochastic model involving two parameters has been conceived  
\cite{biswas-sen1}.
Here a quenched disorder is introduced in  the BS model  through  a parameter $\rho$
representing  the probability that people are completely rigid and never change their opinion
throughout the time evolution. The second  parameter $\mu$  relaxes the rigidity criterion 
in an annealed manner.
It was found that  although with $\rho \neq 0$ and $\mu = 0$, no
consensus state is reached, any nonzero value of $\mu$  
enables the system to reach the all up/down states \cite{biswas-sen1}. 
\subsubsection{The  weighted influence (WI) model}
In all the models described above, the up and down states are taken to be
indistinguishable.  A model in which the up and down domains have 
different weight factors has been considered recently \cite{bss} and 
in fact the 
exit probability was also evaluated as mentioned in the introduction. 
We apply the analysis 
used in this work to the results of \cite{bss} to reveal certain interesting 
features related to the exit probability.
This model was termed as the weighted influence (WI) model, where an individual takes up opinion $1$ with probability
\begin{equation}
P_{1}=\frac{d_{up}}{d_{up} + \delta d_{down}}.
\label{prob1}
\end{equation}
      $\delta$ is the relative influencing ability of the two groups and can vary from zero to $\infty$.
Probability to take opinion value $-1$ is $P_{-1}= 1-P_{1}.$

 All the models discussed above have one common feature in their dynamical rule. For all these models the state of the randomly 
 selected spin depends on the size of the neighbouring domains somehow. In spite of this similarity the intrinsic dynamical rule
 of the models are different. The first one is the BS model which has no disorder, no stochasticity in the dynamics and the 
 state of the selected spin becomes just the same as that of the larger neighbouring domain. Also in the BS model 
 with cutoff there is 
no intrinsic stochasticity and disorder but its late time dynamical behaviour is completely different which is Ising like.
 In the $\beta$ model, thermal noise like disorder is introduced. Here the dynamics is stochastic but for any non-zero 
 $\beta$ the system belongs to the dynamical class of BS model. In the $\rho-\mu$ model disorder is introduced through 
 $\rho$ and $\mu$. For $\mu\ne 0$, this model has the same dynamical behaviour as the BS model. In the $\epsilon$ model, with 
 $\epsilon>0.5$, dynamical behaviour is same as BS model whereas for $\epsilon=0.5$ the dynamics is Ising like and 
 for $\epsilon<0.5$ a different behaviour has been observed previously. On the other hand the WI model, 
 which incorporates stochasticity, 
 is not equivalent to the BS model in any limit. It belongs to a different universality class as far as persistence behaviour is concerned.

System sizes ranging from $L = 200$ to $L = 50000$  were considered depending 
on the model; e.g., since the BS model does not involve any parameter,
one could probe much larger sizes \footnote{In fact to get reliable results for the dynamic exponents in the BS model one needs to simulate 
systems with size at least  ${\mathcal {O}}(10^4)$ \cite{biswas-sen1,biswas-sen2,sen,bsr}.}.
 All the simulations are done for at least $2500$ configurations for each system size.
Random sequential updating is used in all the simulations.

\section{Results }

\subsection{Symmetric models: Exit probability}

We first discuss the results for the behaviour of the 
EP for the models with symmetry where up/down states have the same status. These results show
the existence of an  exponent with a model independent universal value.
\subsubsection{BS model}

\begin{figure} [ht]
\hspace{1.5cm}
\includegraphics[width=10.5cm,angle=0]{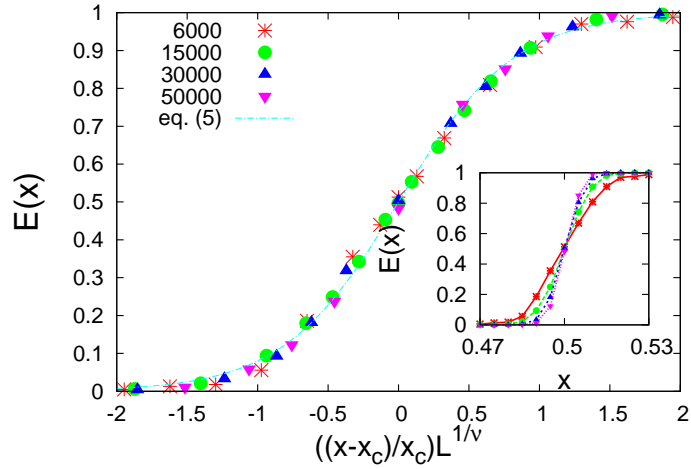}
\caption{ The data collapse of $E(x)$ plotted against $\frac{(x-x_{c})}{x_{c}}L^{1/\nu}$ for different system sizes for BS model. Inset shows the 
unscaled data for the exit probability against initial concentration $x$.}
\label{bsno}
\end{figure}

For the  BS model we have studied the
exit probability $E(x)$ for system sizes ranging from $L=6000$ to $L=50000$.
The plot of $E(x)$ against initial fraction of up spins $x$  shows
that it is nonlinear having strong system size dependence and that the different curves intersect at a single point
 $x_c=0.5 \pm 0.001$ (shown in the inset of Fig. \ref{bsno}). The curves become steeper as the system size is increased. The exit probability thus  shows a step
 function behaviour in the thermodynamic limit. 

Finite size scaling analysis can
 be made using the scaling form
\begin{equation}
 E(x,L)=f\left[\frac{(x-x_{c})}{x_{c}}L^{1/\nu}\right]
\label{scaling}
\end{equation}
where $f(y)\rightarrow0$ for $y<<0$ and equal to $1$ for $y>>0$, so that the data for different 
system sizes $L$ collapse when
 $E(x)$ is plotted against $\frac{(x-x_{c})}{x_{c}}L^{1/\nu}$. The data collapse
takes place with $\nu=2.5\pm0.03$ (Fig. \ref{bsno}). 
The value of $x_c$ has been estimated as follows: ideally at $x_c$ the exit probability is size independent. 
Numerically it is difficult to obtain exact intersection for all the system sizes. 
 Above  $x_c$, data for larger values of system size lie above in the $E(x)$ vs $x$ plane, and below $x_c$ the opposite happens, so 
by observing the range of x for which this happens, the error bars are estimated.  
To obtain $\nu$, one uses the value of $x_c$ obtained as above and the range of $\nu $ values 
for which the collapse appears to be good is taken to estimate the error bars. 
However, a  good scaling collapse can only be obtained for sufficiently large values of $L$, typically  $L > 5000$.

\subsubsection{BS model with cutoff} 

Introducing a cutoff in the BS model, 
     we calculate $E(x)$ using cutoff factor $p > 0$
   for different system sizes ranging from $L=200$ to $L=2000$. As long as $p<1$, the time to reach equilibrium $\sim L^2$. 
   So here we have to restrict system size at $L=2000$ while for other models we use much larger system sizes.
\begin{figure}
\hspace{1.5cm}
\includegraphics[width=11cm,angle=0]{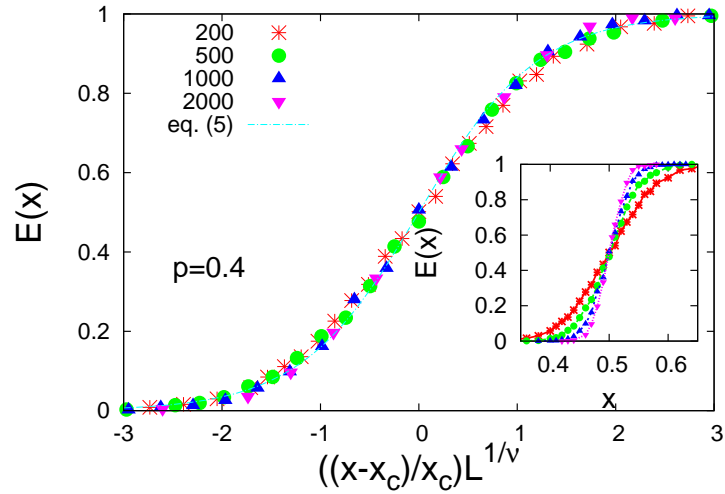}
\caption{The data collapse of $E(x)$ plotted against $\frac{(x-x_{c})}{x_{c}}L^{1/\nu}$  for different system sizes for BS model with cutoff $p=0.4$. Inset shows
the unscaled data for the exit probability against initial concentration $x$. }
\label{cutoff}
\end{figure}

Here $E(x)$ shows (Fig. \ref{cutoff})  scaling behaviour given by eq. (\ref{scaling}) 
   as in the BS model with $\nu=2.5\pm0.03$ and $x_c = 0.5 \pm 0.001$ for any $p$.

\subsubsection{The $\beta$ model}

\begin{figure} [ht]
\hspace{1.5cm}
\includegraphics[width=11cm,angle=0]{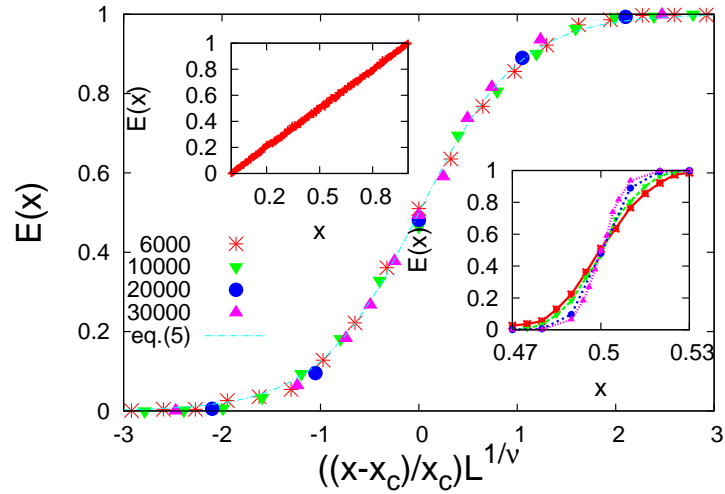}
\caption{ The data collapse of $E(x)$ plotted against $\frac{(x-x_{c})}{x_{c}}L^{1/\nu}$ for different system sizes for model with noise parameter $\beta=0.1$. Inset at the bottom shows the 
unscaled data for the exit 
probability against initial concentration $x$ and inset at the top shows the plot of exit probability for $\beta=0$.}
\label{noise}
\end{figure}

In the $\beta$ model we have studied the exit probability with noise parameter $\beta \geq 0$ 
       for system sizes ranging from $L=6000$ to $L=30000$. For $\beta=0$ the plot of $E(x)$ against $x$ gives a straight line (Fig. \ref{noise})
       which is expected as it is identical to the   nearest neighbour Ising model. In this case the result
       is independent of finite system sizes also. 

       For any nonzero value of $\beta$,  EP shows nonlinear behaviour with strong
       system size dependence (Fig. \ref{noise}) similar to the BS model.
The scaling is once again found to be identical to eq. (\ref{scaling})
with $\nu = 2.5\pm0.03$, independent of $\beta$. Here also 
$x_c =0.5 \pm 0.001 $ for all values of $\beta \neq 0$.

\subsubsection{The $\rho-\mu$ model}
\begin{figure} [ht]
\hspace{1.5cm}
\includegraphics[width=11cm,angle=0]{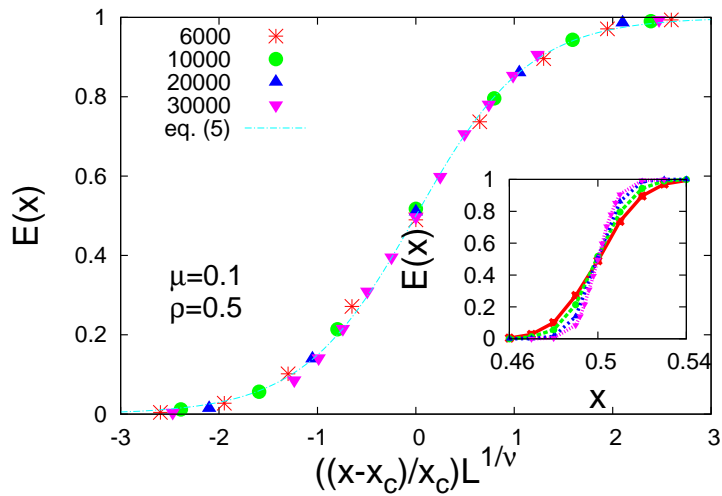}
\caption{ The data collapse of $E(x)$ plotted against $\frac{(x-x_{c})}{x_{c}}L^{1/\nu}$ for $\rho-\mu$ model. Inset shows
the unscaled data for the exit probability against initial concentration $x$.}
\label{rhomufig}
\end{figure}

In the $\rho-\mu$ model we have studied the exit probability with rigidity parameter $\rho\leq1$ for $0< \mu \le 1$
using different system sizes ranging from $L=6000$ to $L=30000$.
For this model also, EP shows a nonlinear behaviour with
 strong system size dependence (Fig. \ref{rhomufig}). Here also $E(x)$ shows  scaling behaviour given by  eq. (\ref{scaling}) with scaling exponent $\nu=2.5 \pm 0.03$ 
independent of $\rho$ and $\mu$. Also $x_c = 0.5 \pm 0.001$ for all values of  $\rho$ and $\mu$.

\subsubsection{The $\epsilon$ model}

For the $\epsilon$ model the system reaches the consensus state (all up/down) only for $\epsilon \geq 0.5$. 
In fact when $\epsilon < 0.5$, the final state is completely disordered.

\begin{center}

\begin{figure} [ht]
\hspace{1.5cm}
\includegraphics[width=11cm,angle=0]{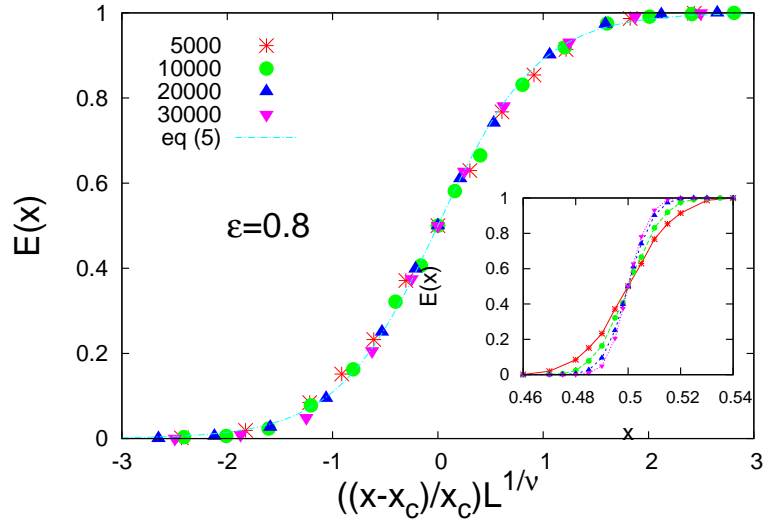}
\caption{ The data collapse  of $E(x)$ plotted against $\frac{(x-x_{c})}{x_{c}}L^{1/\nu}$ for different system sizes for 
$\epsilon$ model with $\epsilon=0.8$. Inset shows the unscaled data for the exit probability 
against initial concentration $x$.}
\label{epsifig}
\end{figure}
 
\end{center}

$E(x)$ is simply equal to $x$ for $\epsilon=0.5$ which is  expected as it is identical to the nearest neighbour Ising model. 
When $\epsilon > 0.5$, EP shows nonlinear behaviour with strong system size dependence 
(Fig. \ref{epsifig} shown for $\epsilon=0.8$) for any value of $\epsilon$. 

The system sizes here  vary  from
$5000$ to $30000$. 
Data collapse similar to the previously discussed models is also obtained here (Fig. \ref{epsifig}) with
$\nu = 2.5 \pm 0.03$ and $x_c = 0.5 \pm 0.001$. 

\subsection{Analysis of the scaling function for symmetric models}

We find that in all the above cases, the EP becomes a step function  at $x = 1/2$ in the thermodynamic limit and 
 the scaling form given by eq. (\ref{scaling}) is obeyed with the value of $\nu=2.5\pm0.03$ being model independent. 
 The scaling function  $f$ is found to fit very well with the general form 
\begin{equation}
f(y) = \left[\tanh (\lambda y) +1\right]/2.
\label{scalingeq}
\end{equation}
 where $ y = \frac{x-x_c}{x_c} L^{1/\nu}$. We conjecture the above form from the following considerations:
first, the shape of the curve suggests a $\tanh$ form (note that the argument $y$ varies from $-\infty$ to $+\infty$).  Secondly, 
 in the thermodynamic limit ($L \rightarrow \infty$), 
$E(x) \rightarrow 0$ for $x < x_c$ and $E(x) \rightarrow 1$ for $x > x_c$ such that one needs to  
add a factor of unity and also a division by 2 in $f(y)$.
This form also leads to the result that   $x_c=1/2$ and  $E(x_c)=1/2$ in the symmetric models, shown
later in this subsection.

\begin{figure} [ht]
\hspace{1.8cm}
\includegraphics[width=11cm,angle=0]{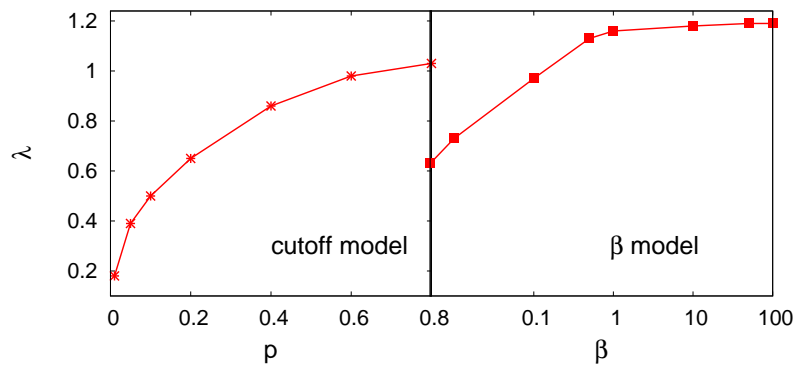}
\caption{ Plot of $\lambda$ against cutoff $p$ (cutoff model; left panel) and against $\beta$ (noise model; right panel)}
\label{cutoff-noise}
\end{figure}

\begin{figure} [ht]
\hspace{1.8cm}
\includegraphics[width=11cm,angle=0]{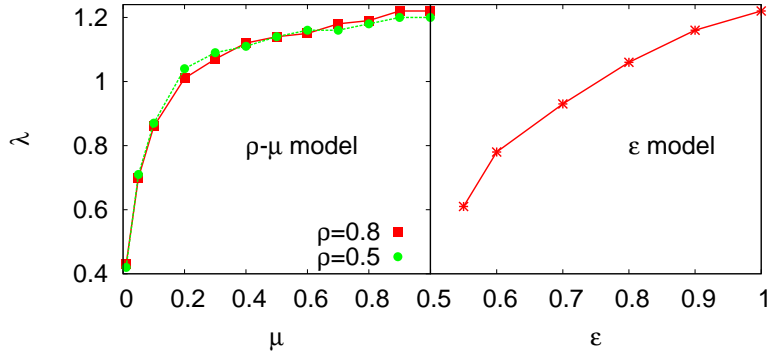}
\caption{ Plot of $\lambda$ against parameter $\mu$ for two fixed values of $\rho=0.5$ and $\rho=0.8$ ($\rho-\mu$ model; left panel) and against parameter 
$\epsilon$ ($\epsilon$ model; right panel)}
\label{epsi-rhomu}
\end{figure}

We obtain the values of $\lambda$ and find that  $\lambda$ is the
factor which is different in each case (Figs \ref{cutoff-noise},  \ref{epsi-rhomu}). 
 For the BS model we found $\lambda = 1.22 \pm 0.02$ by fitting the collapsed plot of the model (fig \ref{bsno}) in 
 equation \ref{scalingeq}. 

When the model involves a parameter, $\lambda$ shows variation  with  the parameter value.

In the cutoff model, $\beta $ model and the $\epsilon$ models, there are limiting values for which the models
coincide with the Ising model and $\lambda$ is undefined. On the other hand, for the $\rho-\mu$ model, 
for $\mu = 0$, $\lambda$ is undefined as the dynamics do not lead to the all up/down state.
As one increases the parameter values beyond that corresponding to the Ising limit in the first three models
mentioned above and $\mu = 0$ in the $\rho-\mu$ model, one finds that $\lambda$ increases.
Each of these four models  becomes equivalent to the BS model in the other extreme  
limiting values of the parameters used. 
Equivalence to the BS model is achieved in the cutoff model  at 
$p=1$; in the $\beta$ model,  for $\beta \to \infty$; in the $\epsilon$ model for $\epsilon = 1.0$ and in the
$\rho-\mu$ model for $\mu =1$ though the BS model behaviour may be present for 
even lesser values of the parameters as far as dynamical exponents are concerned. We find that $\lambda$ varies monotonically and reaches a maximum value 
in the BS limit in general: $\lambda < 1.22$ in the cutoff model, $\epsilon$ model and $\rho-\mu$ model for other values of the parameter 
while in the $\beta$ model, $\lambda$ appears to assume the BS model value beyond a finite value $\beta \approx 1$.

In the cutoff model, a significant change in the timescale occurs as $p \to 1$ \cite{biswas-sen2} and the behaviour of $\lambda $ against $p$ close 
to unity is no longer very smooth. For this reason, we show the results up to $p=0.8$. 
The values of $\lambda$ for the $\beta$ model  shows a rather intriguing  behaviour: it has an increasing behaviour 
for $\beta < 1$ and beyond $\beta=1$,  $\lambda$ increases very slowly and is almost a   constant while approaching the BS value.
However,  there was no perceivable difference observed at $\beta =1$
when other dynamical properties of this model were studied \cite{sen}. 
In the $\rho-\mu$ model, though $\lambda$ depends on $\mu$, we found it to be independent of $\rho$ within error bars. It is due to  the fact that 
$\rho$ is an irrelevant parameter, while $\mu$ is a relevant parameter  as shown in \cite{biswas-sen1}.

In general, one can now use eq.  (\ref{scalingeq}) to write down EP for the symmetric models as 
\begin{equation}
E(x,L) = 
 \frac{\left[\tanh \lambda \left(\frac{x-x_c}{x_c}\right)L^{1/\nu} +1\right]}{2}.
\label{scale_ex} 
\end{equation}
In all these models,  $x_c$ = 1/2 which can be established from symmetry 
arguments. 
In the BS model, 
one has no parameter and $\lambda$ has a unique value.
Using the relation 
\begin{equation}
E(x) + E(1-x) =1, 
\label{exit-identity}
\end{equation}
and putting the expression for $E(x) $ from eq.(\ref{scale_ex}),  one can easily show that $x_c$ has 
to be equal to  $1/2$ and $E(x_c)=1/2$.  
In the other models we  find that $\lambda$ has a dependence on the 
parameter value and in principle one can assume $x_c$ to be a function of the
parameter also, but the observed scaling form and the fact that eq. 
(\ref{exit-identity}) has to be true for all $x$  and $L$ leads to the result
 that $x_c =1/2$ always.

\subsection{Analysis for the asymmetric WI model}
\label{WI}

In the WI model, it had already been noted that the value $\nu \simeq  2.5$ has to be 
used to obtain a  data collapse. The EP   here is found to be of the form
\begin{equation}
E(x,\delta)  = \frac{\left[\tanh \left(\lambda(\delta) y + c(\delta)\right) + 1\right]}{2}
\label{asym}
\end{equation}
where $ y = \frac{x-x_c}{x_c} L^{1/\nu}$; $x_c,  \lambda$ and $c$ all vary with  $\delta$. 

In this model, as the up and down spins have different status, 
it has to be noted  that the  probability that the final state is all down 
starting with $x$ down spins will not be the same as $E(x,\delta)$  for up spins.
Rather, to consider the negative spin case, one has to replace $\delta$ by 
 $1/\delta$  such that 
\begin{equation}
E(1-x,\delta)+E(x,1/\delta) = 1.
\label{WIexit}
\end{equation}
We use the   short-hand notation $x_c$ for $x_c(\delta)$, $c$ for $c(\delta)$ 
and $\lambda$ for $\lambda(\delta)$.   For $1/\delta$, we  use primed 
variables, e.g.  $x_c^{\prime}$ for $x_c(1/\delta)$. 
Putting the expression of $E(x, \delta)$ in eq. (\ref{WIexit}),  we get
\begin{equation*}
 \tanh\left[\lambda \left(\frac{1-x-x_c}{x_c}\right)L^{1/\nu}+c\right]=-\tanh\left[\lambda^{\prime} \left(\frac{x-x_c^{\prime}}{x_c^{\prime}}\right)L^{1/\nu}+c^{\prime}\right].
\end{equation*}
On  simplification one gets 
\begin{equation}
  c+c^{\prime}=L^{1/\nu}\left[x\left(\frac{\lambda}{x_c}-\frac{\lambda^{\prime}}{x_c^{\prime}}\right)-\lambda\left(\frac{1-x_c}{x_c}\right)+\lambda^{\prime}\right].
\label{WIeq}
\end{equation} 

Since $c(\delta)$ and $c(1/\delta)$ cannot have any $L$ dependence then,
\begin{equation}
\label{cdelta}
c(\delta)= - c(1/\delta).
\end{equation}
For the right hand side of (\ref{WIeq}) to be zero for any value of $x$,
\begin{equation}
\label{lambda}
\frac{\lambda(\delta)}{x_c(\delta)}=\frac{\lambda(1/\delta)}{x_c(1/\delta)}
\end{equation}
and,
\begin{equation}
\label{delta}
\lambda(\delta)\left[\frac{1-x_c(\delta)}{x_c(\delta)}\right]=\lambda(1/\delta).
\end{equation}

Eliminating $\lambda(\delta)$ and $\lambda(1/\delta)$ from equations (\ref{lambda}) and (\ref{delta}) we have,
\begin{equation}
x_c(\delta)=1-x_c(1/\delta)
\end{equation}
which  is consistent with the fact that $x$ replaced by $(1-x)$ indicates $\delta$ replaced 
by $1/\delta$ which is the basis of eq. ({\ref{WIexit}).
We have checked that equations (\ref{lambda}) and (\ref{delta}) show excellent 
matching with numerical data. For example for $\delta = 2.0$, we obtain $\lambda=1.6 \pm 0.01$ and $x_c= 0.696 \pm 0.001$ and 
 ( as $1/\delta=1/2=0.5$) the corresponding values for $\delta=0.5$ are $\lambda= 0.7 \pm 0.01$ and $x_c= 0.302 \pm 0.001$. Inserting these values 
we find nearly perfect agreement of both sides of equations (\ref{lambda}) and (\ref{delta}). However, this can be checked 
for any other values of $\delta$ by extrapolating the curve shown 
in the bottom inset of Fig. \ref{lambda-c} and indeed one can get good agreement.

\begin{figure} [ht]
\hspace{1.5cm}
\includegraphics[width=11cm,angle=0]{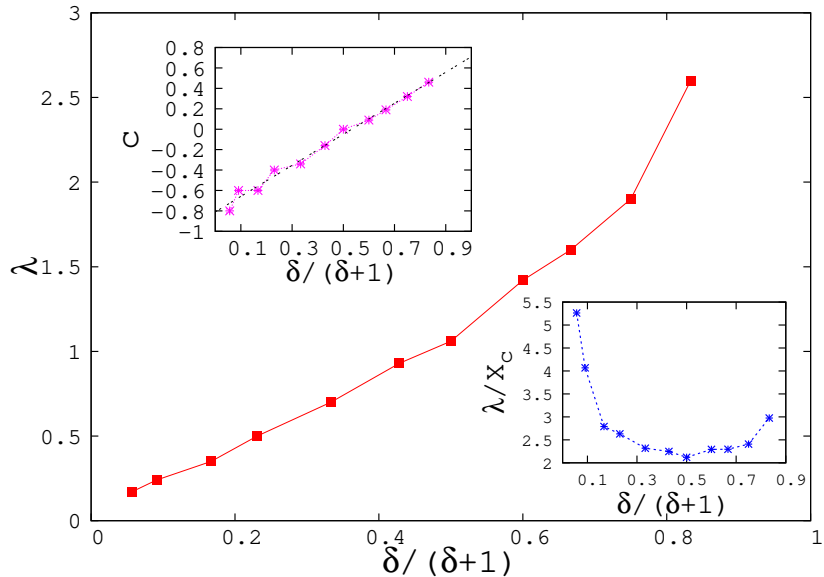}
\caption{ Plot of $\lambda$ against parameter $\delta/(\delta+1)$ and the top inset shows variation of $c$ with parameter $\delta/(\delta+1)$
 and the bottom inset shows variation of $\frac{\lambda}{x_c}$ against $\delta/(\delta+1)$. The dashed straight line at the 
 top inset is the fitted curve based on eq. (\ref{fitting}).}
\label{lambda-c}
\end{figure}

Since the   data for  $c(\delta)$ has a lot of fluctuations for small values of $\delta $ (see inset of Fig.  \ref{lambda-c} ), a direct check of 
 eq. (\ref{cdelta}) is difficult. 
However, from the figure $c(\delta)$ appears to vary linearly with $\delta/(\delta +1)$ and 
one can assume  the form 
\begin{equation}
c(\delta) = a\frac{\delta}{1+\delta} - b.
\label{fitting}
\end{equation}
In case   eq. (\ref{cdelta}) is correct, one must have $b = a/2$.
We tried the above form and the fit appears to be quite accurate with 
$a = 1.53 \pm 0.05$ and $b=0.81 \pm 0.03$ which shows that $b$ is  very close to $a/2$ within error  bar. 
This shows that indeed the scaling form we assumed is consistent with the theory.

In the case of WI model, where $x_c$ is different for different $\delta$, slope of $E(x)$ 
is determined by $\frac{\lambda}{x_c}$ which shows a minimum at $\delta=1$ and increases otherwise (Fig. \ref{lambda-c}, bottom inset).
Thus  any asymmetry makes the EP steeper which is also expected as the asymmetry makes the system
 biased towards one of the absorbing states. 

\section{Summary and discussion}

We have studied a number of opinion dynamics models in one dimension
with different evolutionary rules for the state of the opinions/spins.
A common feature is that the update rules  involve the size of the domains
neighbouring a spin. This immediately gives rise to a different 
behaviour of the exit probability compared to that in the well-studied 
models in 
one dimension.  It shows finite size dependence and 
a step function like behaviour in the thermodynamic limit.
The step function occurs   at $x= 0.5$ for models
in which up and down states carry equal weight.

A  scaling function with a universal form is found to exist with a universal
value of the exponent $\nu$ occurring in it.
Though the scaling function involves a  $\tanh$ term is a conjecture, 
however, we have shown that such a conjecture leads to consistent and meaningful results.
Two non-universal  parameters $\lambda$ and $c$ appear in the scaling function.
$c$ is zero for models which have up/down symmetry. 
$\lambda$ has strong model dependence and it shows interesting variation with the model
parameters. 

The scaling argument for the finite size scaling is $\lambda (\frac{x-x_c}{x_c}) L^{1/\nu}$, which indicates that the width $w$ of the region where $E(x)$ is not equal to unity 
  or zero decreases as $ \frac{x_c}{\lambda}L^{-1/\nu} $.
When the rule is simply a majority rule, i.e.,  the larger neighbouring domain 
dictates the updated state of the spin (BS model),   the value of $\lambda$ is
obtained as $\approx 1.22$.  For the   symmetric models which 
involve a parameter, 
$\lambda$ is smaller than the BS value while  $x_c$ is still equal to 0.5.
This signifies that $w$ is larger for all these models compared to the BS model.
In the WI model where up/down spins are distinguished, 
$\lambda/x_c$ is larger than the BS value as $\delta$ deviates from
unity  showing that $w$ in this case is smaller than the BS value. 
Asymmetry thus plays a strong role in determining the width  $w$.

In the analysis of the WI model, one can further derive equations connecting the values of $x_c$,  $c$ and $\lambda$ at
$\delta$ and $1/\delta$ using eq. (\ref{asym}), which shows very good agreement with numerical 
data. 

We thus arrive at the conclusion that there exists a class of models in one dimension that  shows a behaviour different from familiar short range
spin models in term of EP. Studying different models all of which use a dynamical rule involving the size of the neighbouring domains,  
a universal scaling behaviour accompanied by an exponent with  universal value 
is obtained. The coarsening behaviour of the models considered here are not identical,
 e.g., the cutoff model has a Ising-like late time dynamics (domain growth exponent $z=2$ \cite{biswas-sen2}) while the other models show BS like behaviour
($z=1$). Hence the step function behaviour of EP is clearly due to  the domain size dependent dynamics
as it is known that for the Ising model EP is just a linear function independent of system size. 
 Asymmetry plays an important role but the value of $\nu$ is not affected.

\section{Acknowledgement}
We acknowledge D. Dhar for a critical reading of an earlier version of the manuscript. 
 PR acknowledges financial support from  UGC. PS acknowledges financial support from CSIR project. SB thanks the Department
of Theoretical Physics, TIFR, for the use of its computational resources.

\end{document}